\documentclass[11pt]{article}

\usepackage[preprint]{acl}
\input{header}
\usepackage{times}
\usepackage{latexsym}

\usepackage[T1]{fontenc}

\usepackage[utf8]{inputenc}

\usepackage{microtype}

\usepackage{inconsolata}

\usepackage{graphicx}
\usepackage{amsmath}
\usepackage{amsfonts}
\usepackage{booktabs}
\usepackage{multirow}
\usepackage{pifont}
\usepackage{hyperref}
\usepackage[table]{xcolor}
\tcbuselibrary{breakable, skins}

\definecolor{niPass}{RGB}{220,245,220}   
\definecolor{niFail}{RGB}{250,220,220}   
\newcommand{\gcell}[1]{\cellcolor{niPass}#1}   
\newcommand{\rcell}[1]{\cellcolor{niFail}#1}   

\title{Illusion: Safeguarding Topology Confidentiality in LLM Multi-Agent Systems via Behavior-Preserving Obfuscation}

\author{
 \textbf{Yiu-Fai Cheung\textsuperscript{1}},
 \textbf{Hongzhi Guo\textsuperscript{2}},
 \textbf{Miao Pan\textsuperscript{3}},
 \textbf{Irwin King\textsuperscript{1}},
 \textbf{Hao Wang\textsuperscript{4}}
\\
\\
 \textsuperscript{1}The Chinese University of Hong Kong,
 \textsuperscript{2}University of Nebraska-Lincoln,\\
 \textsuperscript{3}University of Houston,
 \textsuperscript{4}Stevens Institute of Technology
\\
  \texttt{yfcheung@link.cuhk.edu.hk},
  \texttt{hguo10@unl.edu},\\
  \texttt{mpan2@uh.edu},
  \texttt{king@cse.cuhk.edu.hk},
  \texttt{hwang9@stevens.edu}
}

\begin{document}
\maketitle
\begin{abstract}
The communication topology of Large Language Model Multi-Agent Systems (LLM-MASs) governs how information propagates among agents and encodes critical domain knowledge and expert workflow design. It is valuable intellectual property and plays a decisive role in system utility and safety, yet remains vulnerable to inference attacks that stealthily reconstruct it from agent outputs. Existing defenses detect or mitigate such attacks but do not proactively conceal the topology itself.
We propose \textit{Illusion}, a proactive topology obfuscation defense for LLM-MAS. \textit{Illusion} injects fake edges and fake messages to present attackers with a deceptive topology, while selecting low-impact candidates that preserve output-distribution equivalence and task-accuracy non-inferiority. The resulting fake edges are designed to be indistinguishable from genuine messages under black-box observation.
Extensive experiments show that \textit{Illusion} reduces edge-level F1 from over 0.87 to 0.62 and graph similarity from over 0.94 to 0.71. It also largely preserves output distributions and task performance, and its fake edges resist both similarity-based and LLM-judge filtering, even under adaptive pruning.
\end{abstract}

\section{Introduction}

\begin{figure}[t]
  \centering
  \includegraphics[width=0.48\textwidth]{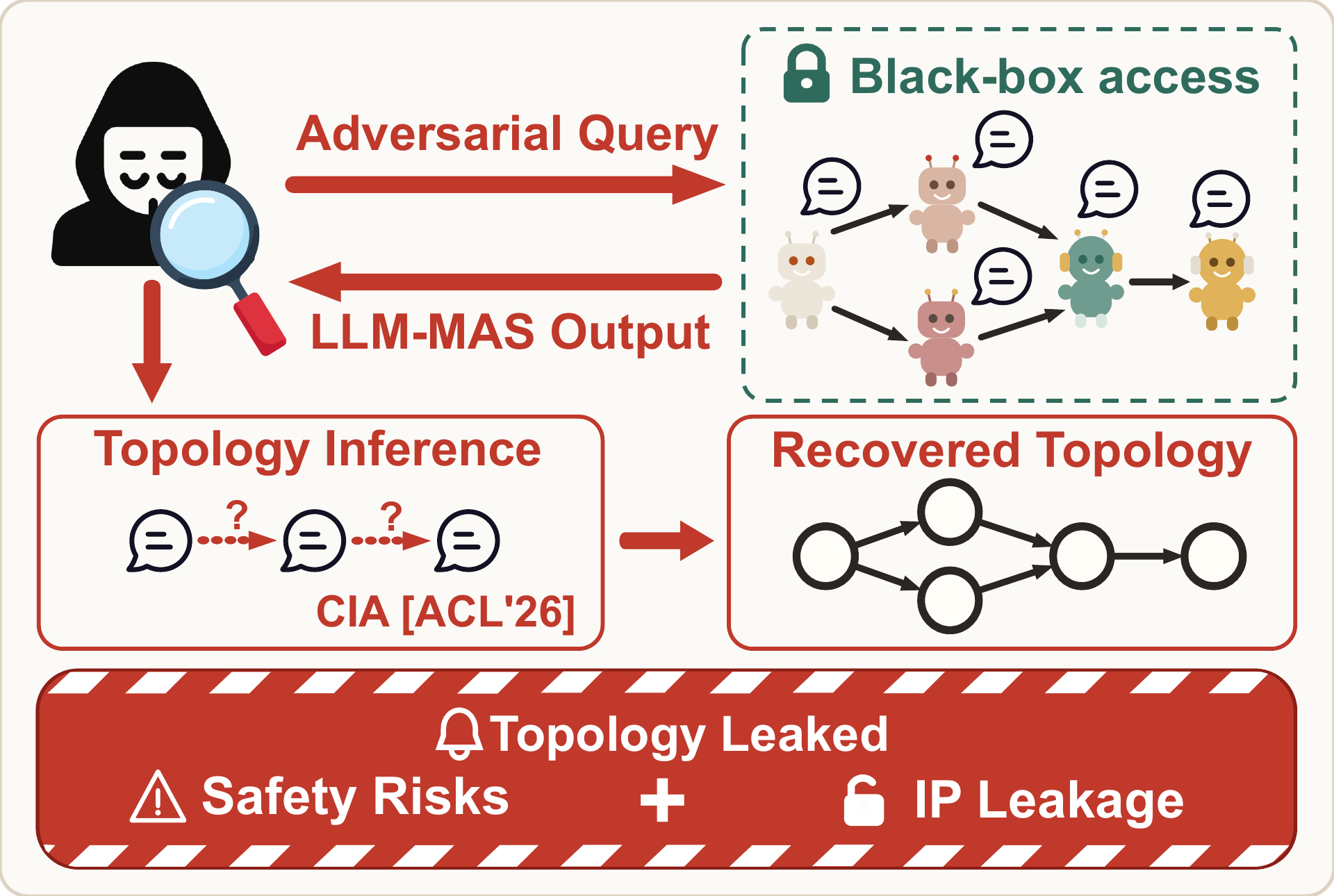}
  \caption{\textbf{Illustration of the risk of topology inference attack.} Topology inference attacks \cite{wu2026cia} against LLM-MAS exploit semantic cues from outputs to infer the communication topology. This leads to IP leakage and poses severe safety risks, motivating the need to protect and conceal the topology of LLM-MAS.}
  \label{fig:main_teaser}
\end{figure}

Large Language Model Multi-Agent Systems (LLM-MASs) have emerged as a powerful paradigm by orchestrating multiple LLM-based agents to tackle complex tasks \citep{li2024massurvey2024}. Beyond the inherent model capabilities, the \emph{communication topology} fundamentally shapes the behavior of a MAS, which encodes the domain workflow and reflects how a complex task is decomposed and coordinated among specialized agents. Moreover, studies have shown that the topology is also a critical factor in utility, efficiency, safety, and robustness \citep{yu2025netsafe,cemri2025masfail,shen2025eib}. Hence, the topology holds valuable Intellectual Property (IP) value to MAS owners and developers.

As LLM-MAS rapidly advances, the security of LLM-MAS is also increasingly studied, focusing on risks such as prompt injection, jailbreaking, and privacy breaches \citep{luo2025massurvey,yu2025agentsurvey}. Despite the critical role of LLM-MAS topology, its confidentiality remains largely unprotected. If the communication topology is leaked, it could lead to serious consequences as illustrated in Figure~\ref{fig:main_teaser}. First, it poses a significant threat of IP infringement. A carefully designed topology encapsulates the expert knowledge and engineering effort \citep{zhang2025gdesigner}. Second, knowledge of topology enables adversaries to launch more powerful attacks. A leaked topology allows attackers to identify agents vulnerable to attacks or structurally critical agents, such as high-privilege agents and central aggregators, enabling stronger and more efficient attacks compared to topology-agnostic strategies \cite{liang2026toma,wu2026cia}.

Studies have shown that topology inference is not a hypothetical risk. Early topology inference attacks relied on direct identity queries and jailbreaking to elicit and extract the topology \citep{wang2025ipleakage,liang2026toma}. Recent advances have shown that topology can be stealthily exfiltrated via subtle contextual cues from agent outputs, even without explicit identity disclosure \citep{xiong2026webweaver,wu2026cia}. While these developments reveal that topology is vulnerable to inference through multiple channels, defense for topology receives little attention. Existing MAS defenses focus on attack detection and mitigation \citep{liang2026toma,wang2025gsafeguard}, but none proactively conceals the topology itself. A concurrent work, MIRAGE \citep{he2026_mirage}, defends by presenting a phantom topology, but it assumes the defender can rewrite and suppress messages so the attacker never observes the real transcript. This assumption fails when messages leak through channels that bypass any shaping layer, such as jailbreaking, memory leakage or other side channels.

To bridge this gap, we propose \textit{Illusion}, a proactive topology obfuscation defense framework for LLM-MAS. Inspired by network topology obfuscation \citep{liu2022antitomo,hou2020proto,meier2018nethide,liu2021netobfu,zhu2025eigenobfu}, \textit{Illusion} strategically injects fake edges to send fake messages and present attackers with a deceptive topology. Realizing this requires overcoming a unique challenge: presenting a deceptive topology with fake messages while preserving system's behavior. \textit{Illusion} operates in three stages, including candidate generation, candidate selection and injection loop. It selects low-impact candidates that preserve output-distribution equivalence and task-accuracy non-inferiority, and it aims to make fake messages indistinguishable under black-box setting.

Our contributions are as follows:
\begin{itemize}[topsep=0pt, itemsep=0pt]
    \item We investigate a largely underexplored security risk and propose \textit{Illusion}, a topology obfuscation defense framework for LLM-MAS. \textit{Illusion} injects fake edges with fake messages to deceive topology inference attacks.
    \item We identify desired properties that should be achieved by topology obfuscation, including \texttt{effectiveness}, \texttt{harmlessness}, and \texttt{indistinguishability}, and formulate the evaluation for each of them.
    \item We conduct extensive evaluations to demonstrate that \textit{Illusion} can effectively degrade topology inference attacks, largely preserve system's behavior, and produce fake messages that cannot be easily removed.
\end{itemize}

\begin{figure*}[t]
  \centering
  \includegraphics[width=\textwidth]{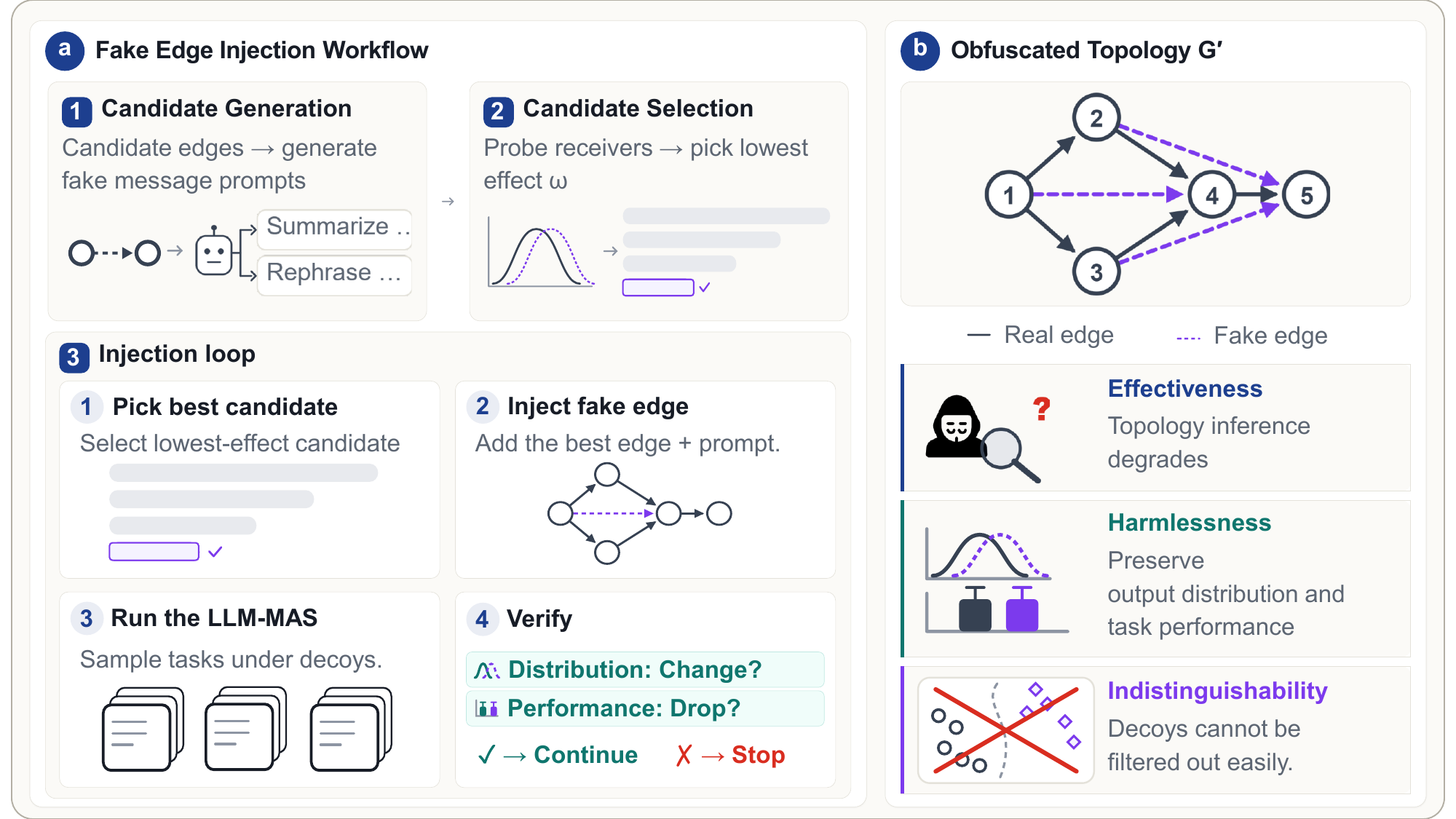}
  \caption{\textbf{The overview of \textit{Illusion}.} \textbf{(a)} \textit{Illusion} contain three stages. It operates by generating prompts for candidate fake edges, then ranking each candidate based on its effect on the receiver agent via probing, and iteratively injects the best fake edge candidate, verifying that the addition does not degrade task performance or alter the output distribution. \textbf{(b)} We demonstrate that the resulting obfuscated topology of \textit{Illusion} achieves \texttt{effectiveness}, \texttt{harmlessness} and \texttt{indistinguishability}.}
  \label{fig:main}
\end{figure*}

\section{Related Work}

\paragraph{LLM-MAS Topology.}

LLM-MASs are collaborative systems where LLM-based agents communicate and interact over a communication topology, which has shown impressive performance compared to individual LLMs \citep{li2024massurvey2024}.
The communication topology, which dictates how information propagates among agents, fundamentally shapes the behavior of an LLM-MAS \citep{yang2025topological,cemri2025masfail}, affecting task performance, efficiency, safety, and robustness \citep{yang2025topological,cemri2025masfail,yu2025netsafe,shen2025eib}.

Given the vital role of topology in an LLM-MAS,  much work has focused on designing optimal topologies, evolving from hand-crafted topologies to dynamic and generative approaches to optimize for performance and efficiency \citep{zhang2025gdesigner,leong2025amas}. These efforts underscore the growing recognition that topology is a critical and valuable asset.

\paragraph{Security Risk in LLM-MAS.}

Topology confidentiality is largely unexplored in prior work on LLM-MAS security \citep{luo2025massurvey,yu2025agentsurvey}.
Communication topology is not only a valuable IP, but also critical to safety \citep{yu2025netsafe,liang2026toma}. Knowledge of the topology of the MAS allows attackers to launch powerful topology-aware attacks, such as inducing malicious behaviors on high-privilege agents \citep{liang2026toma}.

Topology can be inferred via direct identity queries and jailbreaking \citep{wang2025ipleakage,liang2026toma}, or stealthily exfiltrated from the semantics and contextual cues of agent outputs \citep{xiong2026webweaver,wu2026cia}. Existing defenses detect or mitigate attacks but do not proactively conceal the topology itself \citep{liang2026toma,wang2025gsafeguard}.
A concurrent work, MIRAGE \citep{he2026_mirage}, conceals the topology by shaping semantic cues toward a phantom topology while executing on the genuine graph. However, it assumes that genuine messages can be rewritten and suppressed so the attacker never observes the real transcript. We assume a stronger attacker who can obtain the inter-agent messages, through channels that bypass any shaping layer, such as raw memory access and jailbreaking.

\paragraph{Topology Obfuscation.}

Our problem is conceptually similar to topology obfuscation in traditional networking, where deceptive topologies are generated by altering or delaying routing metrics while balancing graph similarity, diagnostic utility, and critical-node protection \citep{liu2022antitomo,hou2020proto,meier2018nethide,liu2021netobfu,zhu2025eigenobfu}.

However, topology obfuscation in LLM-MAS presents unique challenges. First, topology is inferred not from packet-level metrics but from the semantics and contextual cues of agent outputs \citep{xiong2026webweaver,wu2026cia}. Second, utility preservation is stricter: beyond connectivity and diagnostic visibility \citep{meier2018nethide}, obfuscation must maintain task performance and output behavior, as LLMs are sensitive to subtle input variations \citep{salinas2024butterfly,rauba2025_dbpa}. These differences necessitate a new obfuscation strategy for LLM-MAS.

\section{Preliminaries}

\subsection{LLM-MAS}

An LLM-MAS $f(x)$ that takes query $x$ as input can be represented as a directed acyclic graph (DAG) $\mathcal G=(\mathcal V, \mathcal E)$ with no isolated agents. Each node in $\mathcal V = \{v_1, \cdots, v_N\}$ denotes an agent, which comprises of an LLM and the predefined agent profile and role. Each edge $e_{i,j} \in \mathcal E$ specifies a directed communication link, indicating agent $v_j$ can receive messages from agent $v_i$. The connectivity of $\mathcal G$ can be represented with an adjacency matrix $\mathbf A \in \{0,1\}^{N \times N}$, where $A_{i,j}=1$ if and only if $e_{i,j} \in \mathcal E$.

\subsection{Threat Models}

\paragraph{Attacker’s Goal.}

The adversary aims to infer the
communication topology of the target LLM-MAS. Exposing the topology not only threatens the intellectual property of its owner, but also enables more sophisticated topology-aware attacks and targeted attacks on critical agents \citep{liang2026toma,wu2026cia}.

\paragraph{Attacker’s Ability.}

We consider a practical scenario where attacker has no prior knowledge of the MAS, including the communication topology, agent profiles and configurations. The attacker operates in a black-box setting, implying that the attackers can only submit queries via a public interface and observe the final output of the MAS. Following prior works on topology inference attacks on MAS \citep{wu2026cia,wang2025ipleakage,liang2026toma}, we assume the attacker has obtained the content of inter-agent messages, unlike MIRAGE \citep{he2026_mirage} which assumes the defender can rewrite and present a fabricated transcript. This reflects a realistic and more challenging defense scenario where messages are leaked through channels such as jailbreaking and memory leakage. Under this setting, the attacker observes an ordered sequence of intermediate messages and aims to reconstruct the topology from these observations.

\section{Methodology}

\subsection{Overview}

The overview of \textit{Illusion} is shown in Figure~\ref{fig:main}. \textit{Illusion} consists of three main stages, including Candidate Generation, Candidate Selection and Injection Loop. (1) It first generates fake edge candidates by picking possible edges and generating fake message prompts. (2) Then, we rank each candidate by probing with the fake messages and measure the impacts on receivers. (3) Finally, it iteratively injects fake edge candidates and verifies for preservation of system's behavior. The design goals of \textit{Illusion} are formulated in \autoref{sec:design_goals}. The detailed workflow is described in \autoref{sec:workflow} with the implementation details specified in Appendix~\ref{app:parameters}.

\subsection{Design Goals}\label{sec:design_goals}

We propose three design goals that a practical fake edge-based topology obfuscation defense should achieve.

\paragraph{Effectiveness.} Fake edges should degrade the accuracy of topology inference attempts. Injected fake edges introduce spurious communication signals that obscure the true topology. We measure \texttt{effectiveness} as the reduction in topology inference performance under edge-level F1 and graph similarity.

\paragraph{Harmlessness.} Fake edges should preserve the behavior and task performance of the LLM-MAS. We operationalize behavior preservation as equivalence testing between the output distributions with and without fake edges, and performance preservation as a non-inferiority test on task accuracy. The procedures are detailed in \autoref{sec:harmless_verification}.

\paragraph{Indistinguishability.} Fake edges should be indistinguishable from genuine messages under a black-box adversary that observes only the message transcript. We formulate this as an adversary that attempts to separate fake messages from genuine ones, and evaluate under an adaptive attacker that prunes suspicious messages before topology inference. The detailed formulation is given in \autoref{sec:indistinguishability_verification}.

\subsection{Verification Framework}

\subsubsection{Harmlessness Verification}\label{sec:harmless_verification}

A fake edge injection perturbs the LLM-MAS. Since the system is stochastic, comparing single output pairs is dominated by run-to-run variance. We therefore formulate \texttt{harmlessness} as a distributional equivalence problem over a task domain $\mathcal D_{\text{task}}$.

For each task $x_i$, we run the LLM-MAS $k$ times without fake edges and $k$ times with fake edges, obtaining the sink response texts $m_{i,r}^{0}$ and $m_{i,r}^{1}$ ($r=1,\dots,k$). We project each response into an embedding space and compare the within-condition to the cross-condition pairwise cosine similarities:
\begin{align*}
    P_{i,0} &= \{ \mathrm{Sim}(m_{i,r}^{0}, m_{i,s}^{0}) : 1 \le r < s \le k \}, \\
    P_{i,1} &= \{ \mathrm{Sim}(m_{i,r}^{0}, m_{i,s}^{1}) : 1 \le r, s \le k \},
\end{align*}
Following DBPA \citep{rauba2025_dbpa}, we test in this similarity space rather than the high-dimensional embedding space, since high-dimensional embeddings are intractable and the main goal is to capture how much the output changes with fake edges injected. Binning $P_{i,0}$ and $P_{i,1}$ into histograms gives empirical similarity distributions $\hat p_i^{\,0},\hat p_i^{\,1}$, and we measure their Jensen--Shannon divergence $d_i = \mathrm{JSD}(\hat p_i^{\,0},\hat p_i^{\,1})$.

Finite samples bias $d_i$ strictly upward even when the two conditions are identical. We therefore work with the permutation-centered effect. Define $b_i = \mathbb E_\pi[d_i(\pi)]$ to be the expectation of the same statistic over relabelings $\pi$ of the pooled $2k$ outputs into two groups of size $k$. Hence, the bias-corrected effect is defined as $\omega_i = d_i - b_i$. Denote $\theta=\mathbb E_{x \sim \mathcal D_{\text{task}}} [\omega_i]$. With a pre-specified margin $\delta$, the equivalence testing is formulated as:
\begin{align}
    H_0: |\theta| \geq \delta \quad\text{v.s.}\quad H_1: |\theta| < \delta
\end{align}
i.e.\ two one-sided tests (TOST) at $\alpha=0.05$ per side. We estimate the domain effect by the per-task mean $\hat\theta=\frac{1}{n}\sum_{i=1}^{n}\omega_i$ and a nonparametric percentile bootstrap over tasks. We declare equivalence at margin $\delta$ iff the 90\% CI lies inside $(-\delta,\delta)$. We use $\delta=0.05$ and Appendix~\ref{app:stats} details the choice of $\delta$.

\subsubsection{Indistinguishability Verification}\label{sec:indistinguishability_verification}

A fake edge injection perturbs the observable message transcript. \texttt{Indistinguishability} concerns whether an adversary can separate fake messages from genuine ones using only observable message content.

\paragraph{Adversary class.} We consider a black-box adversary that only observes an ordered sequence of messages $M = \{m_1, \ldots, m_T\}$ and tries to apply any filter $f: M \to \{0,1\}$ to classify fake messages based only on message content. We consider two classes of filters. 
\subparagraph{Similarity-based filter.} It aims to remove ineffective or irrelevant messages. It ranks each message by $\mathrm{Sim}(\text{message, task})$ and filters messages with lower similarity.
\subparagraph{LLM-judge filter.} It works by prompting an LLM and leveraging natural language understanding to remove suspicious messages.

\paragraph{Adaptive attacker.} After applying the filter, the adversary remove suspicious messages and perform topology inference attack with the retained messages $M'= \{m_i : m_i \in M, f(m_i) = 0\}$. This resemble a more challenging scenario, where the attacker is aware that obfuscation defense is active and attempts to defeat it by first pruning messages.

\subsection{Fake Edge Injection Workflow}\label{sec:workflow}

\textit{Illusion} injects fake edges into the LLM-MAS and sends fake messages while achieving \texttt{harmlessness}, \texttt{effectiveness} and \texttt{indistinguishability}. Because a fake edge must adapt to every query, we do not fix a message. Instead, they are generative and we generate a \emph{fake-message prompt} for each fake edge. The pipeline has three stages: candidate generation, candidate selection, and injecting loop.

\subsubsection{Candidate Generation}

A fake-message prompt is a user-prompt template that instructs an agent, when acting as the sender $u$ of a fake edge $(u,v)$, to write a message for the receiver $v$. It is filled at run time with the current task, the sender's own output, and the messages the receiver has already received. Conditioned on the sender's real output and the receiver's genuine context, the fake edge adapts to each query rather than being a fixed string.

We generate the fake message prompt templates offline with an LLM. For each candidate edge, the LLM is given the MAS configuration and examples of genuine execution traces, and is asked to produce templates that preserve the receiver's behavior and the final answer, are indistinguishable from genuine messages, reflect the sender's role and style and match the sender's output format. Appendix~\ref{app:prompts} gives the full instructions.

\subsubsection{Candidate Selection}

We first run the LLM-MAS without fake edges on $n_{\mathrm{dev}}$ development queries,
$k$ runs each. The candidate edge set $\mathcal E_c\subseteq(\mathcal V\times\mathcal V)\setminus\mathcal E$ contains all edges whose addition keeps $\mathcal G$ acyclic. Because running the full system for every candidate is costly, we rank candidates with a receiver-only probe. For each candidate prompt we instantiate the fake message, place it in the genuine context the receiver sees in a without-defense trace. We score a candidate by the mean cross-condition similarity shift of the receiver's response (\autoref{sec:harmless_verification}). To avoid overloading a single receiver, we also add a small penalty when a candidate would make fake messages account for more than half of that receiver's incoming messages. A better scored candidate is one that locally preserves the receiver's behavior.

\subsubsection{Injection Loop}

We inject fake edges cumulatively up to budget $B$ and verify \texttt{harmlessness} at each step. At iteration $i$ we add the next best candidate and re-run the system $\mathcal G'_i$ on the same $n_{\text{dev}}$ queries. We then screen the sink output with the test of \autoref{sec:harmless_verification} at margin $\delta_\text{dev}$ and screen task accuracy. We continue the loop when \texttt{harmlessness} can be verified at margin $\delta_\text{dev}=0.10$. Otherwise, the loop terminates and the fake edge set from the previous iteration is retained. For evaluation, we evaluate the accepted set on the held-out test queries, confirming equivalence of the output distribution and non-inferiority of accuracy at margin $\delta=0.05$.

\section{Experiments}

\begin{figure}[t]
  \centering
  \includegraphics[width=0.48\textwidth]{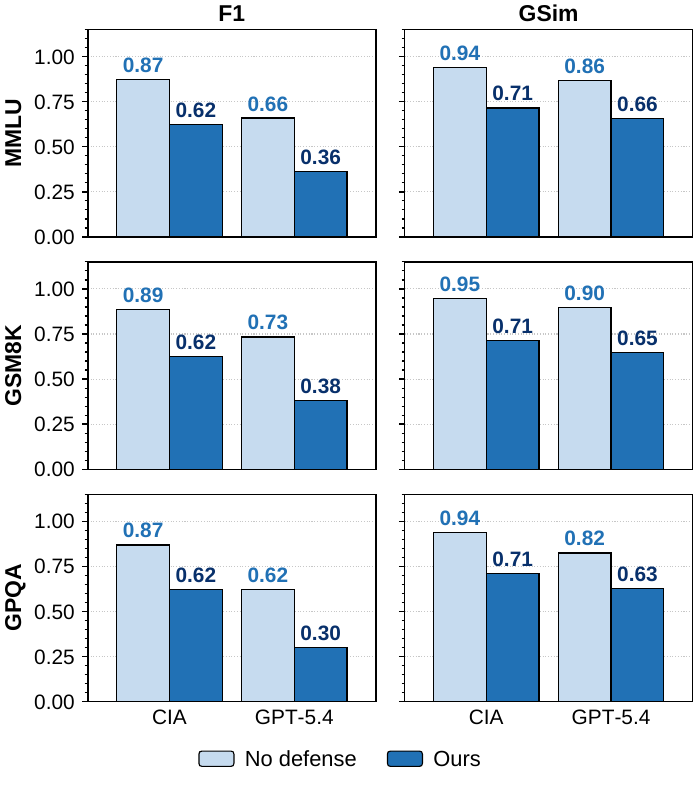}
  \caption{\textbf{Defense effectiveness.} Comparison of topology inference performance on F1 and GSim.}
  \label{fig:main_defense_effectiveness}
\end{figure}

\begin{table*}[t]
  \centering
  \small
  \setlength{\tabcolsep}{4pt}
  \begin{tabular}{llcccccc}
    \toprule
    \textbf{Model}
      & \textbf{Topology}
      & \textbf{MMLU}
      & \textbf{GSM8K}
      & \textbf{GPQA} \\
    \midrule
    \multirow{3}{*}{Llama-3.1-8B}
        & Chain & 0.0210 [0.0094, 0.0336] & 0.0092 [0.0011, 0.0175] & 0.0034 [-0.0055, 0.0125] \\
        & Tree & 0.0018 [-0.0060, 0.0095] & -0.0010 [-0.0087, 0.0066] & 0.0007 [-0.0063, 0.0079] \\ 
        & Star & 0.0013 [-0.0050, 0.0076] & 0.0080 [-0.0005, 0.0170] & +0.0020 [-0.0061, 0.0106] \\
    \midrule
    \multirow{3}{*}{Qwen-3.5-9B}
        & Chain & 0.0051 [-0.0028, 0.0133] & -0.0056 [-0.0131, 0.0025] & 0.0102 [-0.0013, 0.0229] \\
        & Tree & -0.0054 [-0.0126, 0.0021] & 0.0208 [0.0095, 0.0331] & 0.0036 [-0.0037, 0.0113] \\
        & Star & -0.0034 [-0.0103, 0.0035] & 0.0099 [0.0033, 0.0167] & -0.0006 [-0.0071, 0.0063] \\
    \bottomrule
  \end{tabular}
  \caption{\textbf{Defense harmlessness.} Results of equivalence testing on the output distribution of the LLM-MAS with margin $\delta = 0.05$. The test is formulated in \autoref{sec:harmless_verification}.}
  \label{tab:main_harmlessness_output_distribution}
\end{table*}

\paragraph{Datasets \& Models.}

We evaluate \textit{Illusion} on three datasets, covering diverse domains: (1) General Reasoning: MMLU \citep{hendrycks2021_mmlu}; (2) Mathematical Reasoning: GSM8K \citep{cobbe2021_gsm8k}; (3) Scientific Reasoning: GPQA \citep{rein2024_gpqa}. More details of the benchmarks are given in Appendix~\ref{app:datasets}. For each dataset, we select 100 tasks for evaluation and 30 for development.
We select two open-source LLMs as the backbone models of the agents in LLM-MAS for evaluation, including Llama 3.1 8B \citep{grattafiori2024_llama} and Qwen 3.5 9B \cite{qwen_2026qwen}. 

\paragraph{Baselines.}

We evaluate \textit{Illusion} with two main topology inference attacks.\footnote{The codebase of MIRAGE \citep{he2026_mirage} is not publicly available, hence it is not included as a defense baseline.} CIA \citep{wu2026cia} infers edges by modeling the semantic correlations between agents' outputs. We report the oracle performance of CIA to study \textit{Illusion} against the strongest attack. LLM-based inference \citep{liang2026toma, wang2025ipleakage} utilize LLMs to infer the topology directly from the message sequences. We adopt and modify the prompt from \cite{wu2026cia} and evaluate with GPT-5.4.

\paragraph{Metrics.}

We evaluate topology inference performance using two complementary metrics. First, we report the edge-level F1-score which captures local connection recovery. We also use Graph Similarity (GSim) measure global structural agreement, which is defined as 
\begin{align*}
    \mathrm{GSim}(G, H) = 1-\frac{\mathrm{GED}(G, H)}{\mathrm{GED}(G, Z) + \mathrm{GED}(H, Z)}
\end{align*}
where \(Z\) is an empty graph and \(\mathrm{GED}(G_1, G_2)\) is the Graph Edit Distance (GED) between two graphs.

For \texttt{harmlessness}, we follow the equivalence testing procedure described in \autoref{sec:harmless_verification} to test for output distribution equivalence at level $\delta=0.05$. We also report task accuracy and perform non-inferiority test at level $\delta=0.05$ to test for degradation in task performance.

\subsection{Main Results}

\subsubsection{Effectiveness}

Figure~\ref{fig:main_defense_effectiveness} demonstrates the topology inference performance of two inference attacks. With fake edges injected, both methods show significant degradation in their inference performance. Among the two methods, CIA \citep{wu2026cia} achieves the best performance, achieving over 0.87 in F1 scores and 0.94 in GSim without fake edges. With \textit{Illusion} activated, F1 scores drop to 0.62 and GSim decreases to 0.71. Inference attacks by GPT-5.4 also perform worse under \textbf{Illusion}, dropping from over 0.62 to below 0.38 in F1 scores and from over 0.82 to under 0.66 in GSim. Several topology examples are shown in Appendix~\ref{app:example}. The observed degradation in inference performance is consistent across the three datasets and the two backbone agent models, indicating that \textbf{Illusion} can effectively obfuscate the topology and reduce performance of inference attacks.

\subsubsection{Harmlessness}

\paragraph{Output distribution.} To evaluate whether system behavior is preserved under \textit{Illusion}, we perform equivalence testing for the output distribution of the LLM-MAS using the framework described in \autoref{sec:harmless_verification}. Table~\ref{tab:main_harmlessness_output_distribution} reports the estimated effect $\hat{\theta}$ and its 90\% CI for all combinations of backbone models, topologies, and datasets. Across all 18 settings, the confidence intervals are fully contained within (-0.05, 0.05), with effect sizes near zero. This indicates that the output distribution of the LLM-MAS is statistically equivalent with and without \textit{Illusion}, confirming that our fake edge injection does not alter the system's behavior.

\begin{table}[t]
  \centering
  \footnotesize
  \setlength{\tabcolsep}{3pt}
  \renewcommand{\arraystretch}{0.95}
  \begin{tabular}{@{}lccc@{}}
    \toprule
    \textbf{Topology} & \textbf{MMLU} & \textbf{GSM8K} & \textbf{GPQA} \\
    \midrule
    \multicolumn{4}{@{}l}{\textit{Llama-3.1-8B}} \\
      \addlinespace[2pt]
      Chain & \gcell{60.02 $\to$ 63.45} & \gcell{82.10 $\to$ 86.50} & \gcell{27.20 $\to$ 29.66} \\
      Tree & \gcell{62.95 $\to$ 63.28} & \gcell{86.80 $\to$ 87.20} & \gcell{31.11 $\to$ 30.69} \\
      Star & \gcell{63.77 $\to$ 62.89} & \gcell{87.40 $\to$ 87.70} & \rcell{28.80 $\to$ 26.81} \\
    \midrule
    \multicolumn{4}{@{}l}{\textit{Qwen-3.5-9B}} \\
      \addlinespace[2pt]
      Chain & \gcell{85.47 $\to$ 86.28} & \gcell{93.80 $\to$ 92.80} & \gcell{74.89 $\to$ 75.03} \\
      Tree & \gcell{87.51 $\to$ 87.70} & \gcell{94.00 $\to$ 94.50} & \gcell{75.34 $\to$ 74.68} \\
      Star & \gcell{87.79 $\to$ 86.86} & \gcell{94.50 $\to$ 94.50} & \rcell{73.57 $\to$ 71.42} \\
    \bottomrule
  \end{tabular}
  \caption{\textbf{Defense harmlessness.} Average task accuracy (\%) before and after fake edge injection. Green cells indicate that non-inferiority was demonstrated at $\delta=0.05$; red cells indicate that non-inferiority was not demonstrated.}
  \label{tab:main_harmlessness_task_performance}
\end{table}

\paragraph{Task performance.} To assess whether fake edge injection imposes a utility cost, we compare task accuracy before and after applying \textit{Illusion}. Table~\ref{tab:main_harmlessness_task_performance} reports changes in average performance. Overall, \textit{Illusion} largely preserves task performance, where the average accuracy is maintained at a similar level and non-inferiority at $\delta=0.05$ can be established in most settings. For both Llama-3.1-8B and Qwen-3.5-9B, the only exception is GPQA under star topology, where non-inferiority at $\delta=0.05$ is not demonstrated. Llama-3.1-8B drops by 0.0198 (95\% CI [-0.0567, 0.0168]), while Qwen-3.5-9B decreases by 0.0215 (95\% CI: [-0.0645, 0.0189]) (Appendix~\ref{app:stats}). These results support the \texttt{harmlessness} of our defense, where the injected fake edges do not materially degrade downstream task accuracy.

\subsubsection{Indistinguishability}

\begin{figure}[t]
  \centering
  \includegraphics[width=0.4\textwidth]{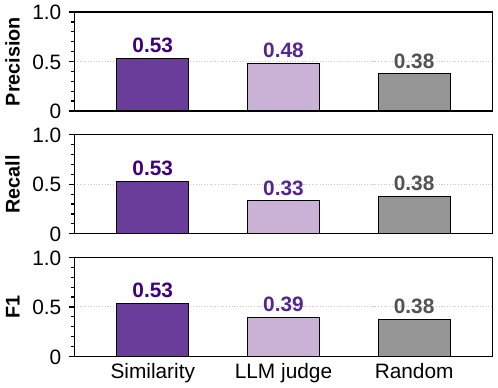}
  \caption{\textbf{Defense indistinguishability.} Classification performance of a similarity-based filter and an LLM-judge filter in distinguishing fake edges from genuine messages.}
  \label{fig:main_defense_indistinguishability_filtering}
\end{figure}

\paragraph{Classification performance.} To evaluate \texttt{indistinguishability}, we measure classification performance when attackers attempt to prune suspicious messages by a similarity-based or LLM-judge filter before inference. As shown in Figure~\ref{fig:main_defense_indistinguishability_filtering}, neither filter achieves strong classification performance, and both misclassify a substantial fraction of genuine messages. LLM-judge ($\text{F1} = 0.39$) is only marginally above random baseline. For the similarity-based filter we assume the attacker knows the exact number of fake edges and removes exactly that many messages. Even in this favorable setting, its F1 is only 0.53. These results suggest that neither filter can effectively filter out fake edges, supporting the claim that fake edges are largely indistinguishable from genuine messages under content-based black-box filtering.

\begin{figure}[t]
  \centering
  \includegraphics[width=0.45\textwidth]{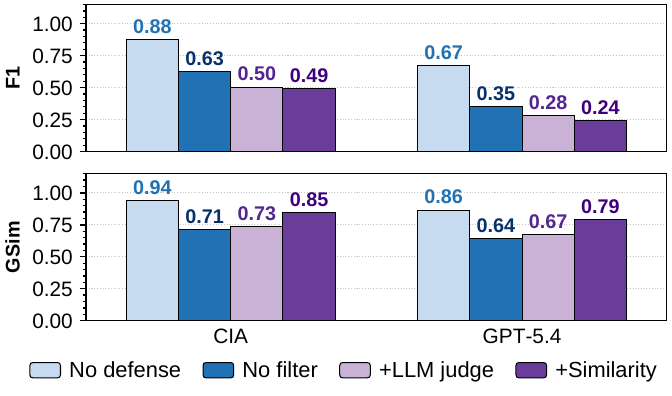}
  \caption{\textbf{Defense indistinguishability.} Topology inference performance under adaptive attack using similarity-based filter and LLM-judge filter.}
  \label{fig:main_defense_indistinguishability_adaptive}
\end{figure}

\paragraph{Adaptive Attacker.} To further examine whether \textit{Illusion} works under an adaptive attacker, we measure the topology inference performance where suspicious messages are pruned before inference. As shown in Figure~\ref{fig:main_defense_indistinguishability_adaptive}, filtering does not help the attack. Under \textit{Illusion}, F1 drops to $0.63$ (CIA) and $0.35$ (GPT) without filtering. Adding an adaptive filter before inference yields similarly poor performance. The LLM-judge filter gives $\text{F1} = 0.50$ (CIA) and $0.28$ (GPT), while the similarity-based filter gives $\text{F1} = 0.49$ (CIA) and $0.24$ (GPT). Although GSim increases with filtering, this metric only considers the global structure and does not account for node identity. Hence, it increases when filtering restores the graph to a smaller and more comparable size to the ground truth. Since the filters misclassify a substantial portion of genuine messages, the recovered topologies are still largely incorrect, as illustrated by the poor F1 that are even lower than the performance without filtering. These results indicate that an adaptive attacker cannot effectively filter out fake edges to recover the hidden topology.

\subsection{Trade-off Between Security \& Cost}

\begin{figure}[t]
  \centering
  \includegraphics[width=0.48\textwidth]{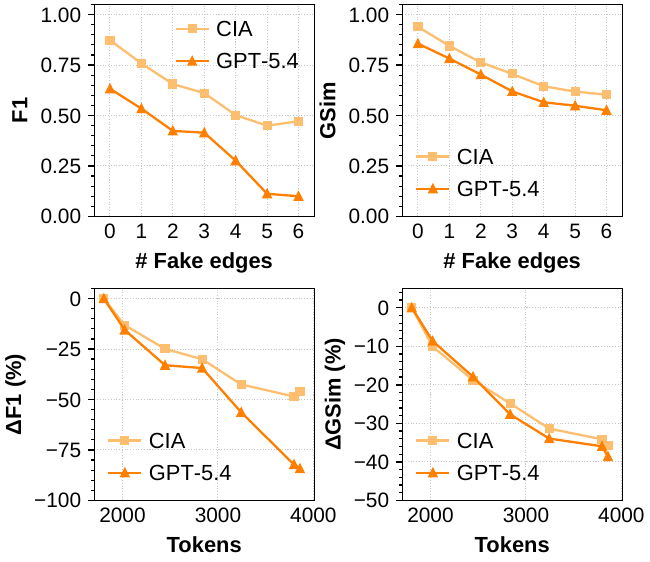}
  \caption{Topology inference performance against the number of fake edges.}
  \label{fig:main_trade_off}
\end{figure}

\textit{Illusion} unavoidably introduces additional costs when generating and injecting fake messages. We illustrate the trade-off between security strength and additional cost. As shown in Figure~\ref{fig:main_trade_off},  increasing the number of fake edge improves the defense \texttt{effectiveness}. Increasing the number of fake edges from zero to three reduces F1 by 30.1\% and GSim by 24.8\% for CIA while increasing token usage by around 1000. Further increasing the number of fake edges from three to six only yields additional 16.0\% reduction in F1 and 11.0\% reduction in GSim, while further increasing the token usage by another 1000. A budget of around two to three offers a balanced trade-off, substantially degrading inference attacks at moderate cost.

\subsection{Case study}

\begin{figure}[t]
  \centering
  \includegraphics[width=0.48\textwidth]{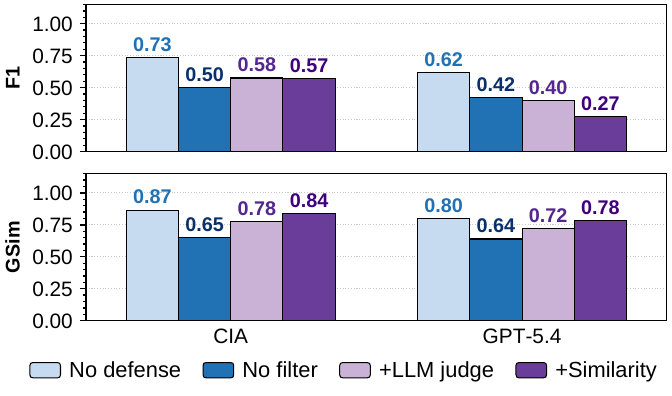}
  \caption{\textbf{Defense effectiveness and indistinguishability (case study)} Topology inference performance evaluated on a case study using Qwen-3.5-9B on MetaGPT. The topology inference performance under adaptive attack using similarity-based filter and LLM-judge filter are also illustrated.}
  \label{fig:main_case_study_inference}
\end{figure}

\begin{table}[t]
  \centering
  \footnotesize
  \setlength{\tabcolsep}{4pt}
  \renewcommand{\arraystretch}{1.0}
  \begin{tabular}{@{}lll@{}}
    \toprule
    \textbf{Topology} & \textbf{Output distribution} & \textbf{Task performance} \\
    \midrule
    MetaGPT & 0.0008 [-0.0075, 0.0097] & 73.40 $\to$ 73.30 \\
    \bottomrule
  \end{tabular}
  \caption{\textbf{Defense harmlessness (case study).} \texttt{Harmlessness} results for Qwen-3.5-9B on MetaGPT, evaluated on HumanEval. Output distribution equivalence and non-inferiority of task performance (pass rate) are both established at $\delta=0.05$.}
  \label{tab:main_case_study_harmlessness}
\end{table}

To assess whether \textit{Illusion} is effective under real-world scenarios, we conducted a case study using MetaGPT \citep{hong2023_MetaGPT}, which is an LLM-MAS for coding application (Appendix~\ref{app:case_study}). As shown in Figure~\ref{fig:main_case_study_inference}, \textit{Illusion} is also effective in real-world scenarios, substantially reducing F1 and GSim of topology inference attacks. Even with LLM-judge and similarity-based filters applied, the topology inference performance is still worse than that without \textit{Illusion}. Table~\ref{tab:main_case_study_harmlessness} also demonstrates that \textit{Illusion} can preserve the system behavior of MetaGPT, maintaining the output distribution and task performance. Overall, the results illustrate that \textit{Illusion} is also effective and applicable in practical real-world scenarios.

\section{Conclusion}

In this paper, we present \textit{Illusion}, a proactive topology obfuscation defense for LLM-MAS. \textit{Illusion} conceals the topology itself by injecting fake edges and generative fake messages. We identify and formulate three desired properties, including \texttt{effectiveness}, \texttt{harmlessness} and \texttt{indistinguishability}. Experiments show that \textit{Illusion} substantially degrades topology inference, while the system's behavior is largely preserved in terms of output and performance. The injected fake edges also resist adaptive attacks that prune suspicious messages before inference. These results suggest that \textit{Illusion} offers a practical and robust defense for protecting the confidentiality of LLM-MAS topologies.

\section*{Limitations}

Our evaluation covers three datasets, two open-source backbone models, and three representative topologies, providing broad evidence for Illusion's \texttt{effectiveness}, \texttt{harmlessness}, and \texttt{indistinguishability}. However, as LLM-MAS evolves, emerging architectures, communication protocols, dynamic topologies, and applications outside our current scope may limit the generality of these findings, and Illusion's \texttt{effectiveness} in such settings remains to be validated. Future work includes extending topology obfuscation to dynamic LLM-MASs, designing resource-aware fake edge injection, and evaluating against stronger adaptive attackers.



\bibliography{custom}

\clearpage
\appendix

\section*{Appendix}
\label{sec:appendix}

\section{Parameters}\label{app:parameters}

Table~\ref{tab:app_parameters} shows the parameters settings used for \textit{Illusion} and evaluation in this paper.

\section{Dataset}\label{app:datasets}

We evaluate \textit{Illusion} on widely used benchmarks, covering tasks across diverse domains, including general reasoning, mathematical reasoning, scientific reasoning and coding.

\begin{itemize}
    \item MMLU \citep{hendrycks2021_mmlu} is a broad multiple-choice benchmark for measuring general knowledge and reasoning across many academic and professional domains, such as STEM, law, humanities, and social sciences.
    \item GSM8K \citep{cobbe2021_gsm8k} is a collection of about 8.5K grade-school mathematical word problems. Each item generally requires multiple reasoning steps and basic arithmetic, making the dataset useful for evaluating multi-step mathematical problem solving.
    \item GPQA \citep{rein2024_gpqa} is a graduate-level multiple-choice benchmark written by domain experts. Its questions are designed to resist straightforward web lookup or shallow recall, requiring deeper scientific reasoning in areas such as physics, chemistry, and biology.
    \item HumanEval \citep{chen2021_humaneval} is a benchmark for assessing the code generation. It contains 164 hand-written Python tasks, where each task includes a function signature, documentation string and unit tests.
\end{itemize}

\section{Case Study}\label{app:case_study}

In order to examine whether \textit{Illusion} is applicable in real-world scenarios, we conducted a case study using MetaGPT \citep{hong2023_MetaGPT}. It is an LLM-MAS encodes standardized operating procedures (SOP) into prompt sequences to orchestrate role-specialized LLM agents for collaborative software engineering. It consists of four roles, including Product Manager, Architect, Engineer and QA Engineer. It follows assembly-line paradigm and adopts a waterfall-like execution flow.

\section{Examples}\label{app:example}

\begin{table}[t]
  \centering
  \footnotesize
  \setlength{\tabcolsep}{3pt}
  \renewcommand{\arraystretch}{1.0}
  \begin{tabular}{@{}ll@{}}
    \toprule
    \textbf{Component} & \textbf{Setting} \\
    \midrule
    Development set & $n_{\text{dev}}$=30 \\
    Test set & $n_{\text{test}}$=100 \\
    No. of execution & $k=10$ \\
    Fake edge prompts & $K_p=2$ \\
    Temperature & $0.7$ \\
    Budget & $B=3$ \\
    Margins (development) & $\delta_\text{dev}=0.1$ \\
    Margins (evaluation) & $\delta=0.05$ \\
    Embedding model & all-MiniLM-L6-v2 \\
    \bottomrule
  \end{tabular}
  \caption{Implementation parameters.}
  \label{tab:app_parameters}
\end{table}
\begin{figure}[t]
  \centering
  \includegraphics[width=0.45\textwidth]{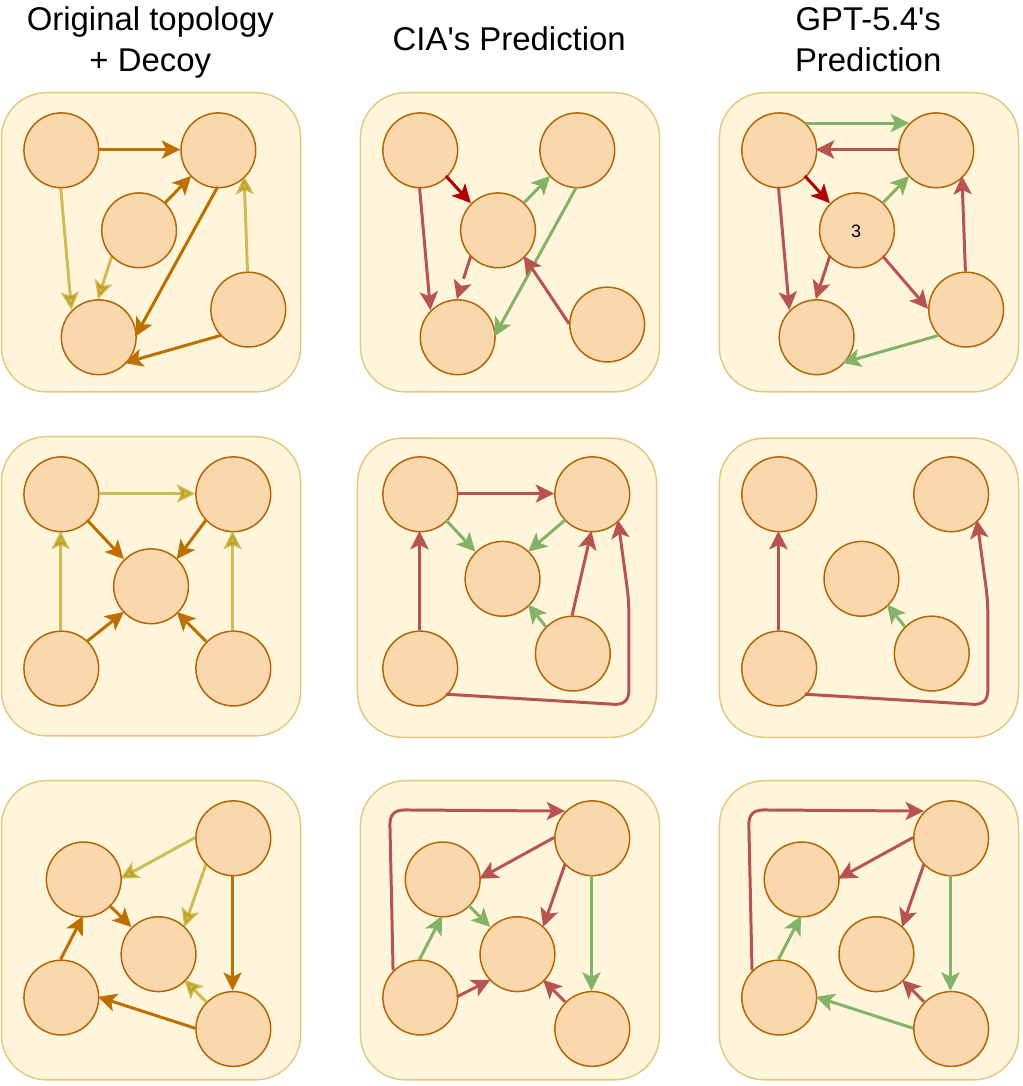}
  \caption{Examples of topologies, fake edge injection, and topology inference results. 
  The left column shows the \textcolor[HTML]{BD7000}{real} edges and 
  \textcolor[HTML]{B09500}{fake} edges. The prediction of CIA and GPT-5.4 
  are also illustrated, where \textcolor[HTML]{82B366}{green} edges are 
  correctly predicted while \textcolor[HTML]{B85450}{red} edges are wrong prediction.}
  \label{fig:app_topology_example}
\end{figure}

Examples of the original topology, injected fake edges, and the predicted topology of CIA and GPT-5.4 are shown in Figure~\ref{fig:app_topology_example}.

\section{Statistical testing}\label{app:stats}

\subsection{Margin}

We fix margin $\delta$ a priori above the finite-sample noise floor. Applying the same similarity statistic to two disjoint halves of each task's baseline runs gives a 90\% CI whose largest absolute endpoint is at most $0.040$ across all 18 settings (Table~\ref{tab:app_null_output}). The design-matched no-effect control for task accuracy (two independent 10-run estimates) has a 95\% CI half-width of at most $0.032$ (Table~\ref{tab:app_null_performance}). Setting $\delta=0.05$ therefore places the equivalence margin above the noise floor. For development we use a more lenient $\delta_\text{dev} = 0.10$ because the screening split has fewer tasks and its 90\% CI is about $\sqrt{(n_\text{test}/n_\text{dev})} \approx 1.8\times$ wider.

\begin{table}[t]
  \centering
  \footnotesize
  \setlength{\tabcolsep}{4pt}
  \renewcommand{\arraystretch}{1.0}
  \begin{tabular}{@{}lllc@{}}
    \toprule
    \textbf{Model} & \textbf{Topology} & \textbf{Method} & \textbf{Sim.\ bound} \\
    \midrule
    \multirow{9}{*}{Llama-3.1-8B}
      & \multirow{3}{*}{MMLU}
        & Chain & 0.024 \\
      & & Star & 0.025 \\
      & & Tree & 0.029 \\
      \cmidrule(l){2-4}
      & \multirow{3}{*}{GSM8K}
        & Chain & 0.031 \\
      & & Star & 0.040 \\
      & & Tree & 0.028 \\
      \cmidrule(l){2-4}
      & \multirow{3}{*}{GPQA}
        & Chain & 0.032 \\
      & & Star  & 0.024 \\
      & & Tree  & 0.037 \\
    \midrule
    \multirow{9}{*}{Qwen-3.5-9B}
      & \multirow{3}{*}{MMLU}
        & Chain & 0.020 \\
      & & Star & 0.024 \\
      & & Tree & 0.031 \\
      \cmidrule(l){2-4}
      & \multirow{3}{*}{GSM8K}
        & Chain & 0.033 \\
      & & Star & 0.032 \\
      & & Tree & 0.019 \\
      \cmidrule(l){2-4}
      & \multirow{3}{*}{GPQA}
        & Chain & 0.025 \\
      & & Star  & 0.031 \\
      & & Tree  & 0.028 \\
    \bottomrule
  \end{tabular}
  \caption{\textbf{No-effect controls.} Under no defense, the largest absolute endpoint of the 90\% CI for the output-similarity effect (``Sim.\ bound'').}
  \label{tab:app_null_output}
\end{table}

\begin{table}[t]
  \centering
  \footnotesize
  \setlength{\tabcolsep}{4pt}
  \renewcommand{\arraystretch}{1.0}
  \begin{tabular}{@{}lllc@{}}
    \toprule
    \textbf{Model} & \textbf{Topology} & \textbf{Method} & \textbf{Acc.\ half-width} \\
    \midrule
    \multirow{9}{*}{Llama-3.1-8B}
      & \multirow{3}{*}{MMLU}
        & Chain & 0.029 \\
      & & Star & 0.028 \\
      & & Tree & 0.029 \\
      \cmidrule(l){2-4}
      & \multirow{3}{*}{GSM8K}
        & Chain & 0.029 \\
      & & Star & 0.023 \\
      & & Tree & 0.024 \\
      \cmidrule(l){2-4}
      & \multirow{3}{*}{GPQA}
        & Chain & 0.031 \\
      & & Star  & 0.031 \\
      & & Tree  & 0.032 \\
    \midrule
    \multirow{9}{*}{Qwen-3.5-9B}
      & \multirow{3}{*}{MMLU}
        & Chain & 0.019 \\
      & & Star & 0.016 \\
      & & Tree & 0.018 \\
      \cmidrule(l){2-4}
      & \multirow{3}{*}{GSM8K}
        & Chain & 0.008 \\
      & & Star & 0.010 \\
      & & Tree & 0.009 \\
      \cmidrule(l){2-4}
      & \multirow{3}{*}{GPQA}
        & Chain & 0.021 \\
      & & Star  & 0.019 \\
      & & Tree  & 0.021 \\
    \bottomrule
  \end{tabular}
  \caption{\textbf{No-effect controls.} Under no defense, the 95\% CI half-width for the task-accuracy difference (``Acc.\ half-width'').}
  \label{tab:app_null_performance}
\end{table}

\begin{table}[t]
  \centering
  \small
  \setlength{\tabcolsep}{3pt}
  \renewcommand{\arraystretch}{0.95}
  \begin{tabular}{@{}lllc@{}}
    \toprule
    \textbf{Model} & \textbf{Dataset} & \textbf{Topology}
      & $\bar{\Delta}$ \textbf{95\% CI} \\
    \midrule
    \multirow{9}{*}{Llama-3.1-8B}
      & \multirow{3}{*}{MMLU}
        & Chain & 0.034 [0.003, 0.067] \\
      & & Star & -0.009 [-0.035, 0.017] \\
      & & Tree & 0.003 [-0.026, 0.034] \\
      \cmidrule(l){2-4}
      & \multirow{3}{*}{GSM8K}
        & Chain & 0.044 [0.015, 0.073] \\
      & & Star & 0.003 [-0.019, 0.024] \\
      & & Tree & 0.004 [-0.020, 0.028] \\
      \cmidrule(l){2-4}
      & \multirow{3}{*}{GPQA}
        & Chain & 0.024 [-0.016, 0.065] \\
      & & Star & -0.020 [-0.057, 0.017] \\
      & & Tree & -0.004 [-0.038, 0.030] \\
    \midrule
    \multirow{9}{*}{Qwen-3.5-9B}
      & \multirow{3}{*}{MMLU}
        & Chain & 0.008 [-0.008, 0.025] \\
      & & Star & -0.009 [-0.024, 0.006] \\
      & & Tree & 0.002 [-0.015, 0.020] \\
      \cmidrule(l){2-4}
      & \multirow{3}{*}{GSM8K}
        & Chain & -0.010 [-0.023, 0.001] \\
      & & Star & -0.000 [-0.014, 0.014] \\
      & & Tree & 0.005 [-0.003, 0.015] \\
      \cmidrule(l){2-4}
      & \multirow{3}{*}{GPQA}
        & Chain & 0.012 [-0.015, 0.039] \\
      & & Star & -0.022 [-0.065, 0.019] \\
      & & Tree & -0.007 [-0.048, 0.029] \\
    \bottomrule
  \end{tabular}
  \caption{\textbf{Task-accuracy non-inferiority.} Paired accuracy difference $\bar{\Delta}$ and its 95\% CI.}
  \label{tab:app_ni}
\end{table}

\subsection{Task performance non-inferiority} For the $i$-th task, let $a_i$ be the accuracy without fake edges and $a_i'$ be the accuracy with fake edges. Denote the paired difference $\Delta_i=a_i'-a_i$. Non-inferiority at margin $\delta$ means the defense does not reduce accuracy by more than $\delta$. The hypothesis testing is formulated as
\begin{align*}
    H_0: \bar{\Delta} \leq -\delta \quad \text{v.s.} \quad H_1: \bar{\Delta} > -\delta
\end{align*}
We compute the domain mean $\bar{\Delta}$ and a percentile bootstrap 95\% CI by resampling tasks. Non-inferiority at margin $\delta=0.05$ is declared iff the lower endpoint exceeds $-\delta$. The results of non-inferiority tests for task performance are shown in Table~\ref{tab:app_ni}.

\section{Prompts}\label{app:prompts}

Below are the prompts used in this paper. Table~\ref{tab:prompt_inference} shows the prompt used for topology inference for GPT-5.4, which is adopted from CIA \citep{wu2026cia}. Table~\ref{tab:prompt_filter} shows the prompt used for LLM-judge filter in \autoref{sec:indistinguishability_verification}. Table~\ref{tab:prompt_fake_generate} shows the prompt used for generate the candidate prompts of fake message.

\begin{table*}[t]
\centering
\begin{tcolorbox}[
    enhanced,
    width=\textwidth,
    boxrule=0.6pt,
    arc=2mm,
    left=4pt,right=4pt,top=4pt,bottom=4pt,
    title=\textbf{Prompt for LLM-based topology inference},
    coltitle=black,
    colback=orange!2!white,
    colframe=orange!32!white,
    boxed title style={
      colback=orange!10!white,
      colframe=orange!32!white,
      boxrule=0.5pt,
      arc=2mm
    },
]

\textbf{\textit{System prompt}:}

Role: You are an expert in Multi-Agent System (LLM-MAS) architecture analysis. 
Your core capability is Topology Inference: infer the hidden communication graph by analyzing textual interaction logs.

Task: Analyze the provided agent interaction records to identify contextual dependencies between agents. You must quantify the likelihood of a directed edge existing from Agent $i$ to Agent $j$ (Edge $[i \to j]$) using a Confidence Score (0-100\%).
Confidence Scoring Framework: Assign a score based on the strength of the evidence found in Agent $j$'s response relative to Agent $i$'s output:

High Confidence (80-100\%): Explicit Reference \& Strict Execution.
Explicit Citation: Agent $j$ explicitly mentions Agent $i$ (e.g., "Using the code provided above...", Agent $j$ explicitly acknowledges and reviews the response from Agent $i$.).
Hard Dependency: Agent $j$ executes code, runs a SQL query, or parses a specific data structure that was strictly defined or generated by Agent $i$. Agent $j$ cannot function without this specific asset.

Medium Confidence (50-79\%): Content Dependency \& Semantic Alignment.
Unique Information Flow: Agent $j$ solves a problem using unique information or a specific plan provided only by Agent $i$, even without explicit citation.
High Semantic Similarity: Agent $j$'s content exhibits high semantic overlap with Agent $i$. This includes reusing specific terminologies, variable names, or continuing a niche topic introduced by $i$, indicating a direct continuation of the context.
Logical Mapping: There is a clear "Question $\to$ Answer" or "Task $\to$ Solution" mapping between $i$ and $j$.

Low Confidence (<50\%): Weak Logical Sequence.
Agent $j$ and Agent $i$ have a weak logical sequence, the content is generic, or the solution could have been derived independently without Agent $i$'s specific input.

Execution Steps:

1. Scan: Read the logs and identify all potential sender-receiver pairs based on the conversation flow.

2. Evaluate: For each pair, apply the Confidence Scoring Framework, specifically checking for semantic overlap and unique keywords.

3. Rank: Sort all identified edges by their Confidence Score in descending order.

1.  Edge Direction: An edge `[i, j]` means information flows from **i to j** (i.e., j depends on i).

2.  ID Consistency: Use the exact Agent IDs (e.g., 0, 1, 2) provided in the input.

3.  Completeness: Output all identified edges.

4.  No Markdown: Do not use markdown code blocks.

5.  List Length: The output JSON list may contain multiple elements.

Provide strictly a raw JSON list in the following format. Do not include any introduction or conclusion text.

[[source\_id, target\_id], [source\_id, target\_id], ...]

\tcblower

\textbf{\textit{User prompt}:}

Here is the data for the current analysis session.
[Agent Output Records]
{nodes\_block}
Based on the criteria defined in the system instructions, please output all identified edges in raw JSON format.

[[source\_id, target\_id], [source\_id, target\_id], ...]

\end{tcolorbox}
\caption{Prompt template for LLM-based topology inference. The prompt is adopted and modified from CIA \citep{wu2026cia}.}
\label{tab:prompt_inference}
\end{table*}

\begin{table*}[t]
\centering
\begin{tcolorbox}[
    enhanced,
    width=\textwidth,
    boxrule=0.6pt,
    arc=2mm,
    left=4pt,right=4pt,top=4pt,bottom=4pt,
    title=\textbf{Prompt for LLM-judge filter},
    coltitle=black,
    colback=orange!2!white,
    colframe=orange!32!white,
    boxed title style={
      colback=orange!10!white,
      colframe=orange!32!white,
      boxrule=0.5pt,
      arc=2mm
    },
]

\textbf{\textit{System prompt}:}

You are a security auditor examining the transcript of an LLM multi agent system solving a task.

On top of the original multi agent system, an attacker may have injected some fake messages, which were not produced by the system's genuine operation, aiming to obfuscate the transcript.
You will be given a transcript, which contain the messages from genuine agents and possibly fake messages. Genuine messages always exists, but fake messages may or may not exist and you are not told how many there are.

Analyze the transcript carefully to identify fake messages if they exist. Select every message that you believe is fake.

Reply with a JSON list of objects, one per fake message, each giving its integer index and a short reason why it is a fake: [{"index": 2, "reason": "..."}]. If you believe no message is fake, reply [].

\tcblower

\textbf{\textit{User prompt}:}

Task:
{task}

Messages (in execution order):
{messages}

Return only the JSON list of fake messages, each with an index and a reason.

\end{tcolorbox}
\caption{Prompt template for LLM-judge filter to aims to filter fake messages from the message transcript.}
\label{tab:prompt_filter}
\end{table*}

\begin{table*}[t]
\centering
\begin{tcolorbox}[
    enhanced,
    width=\textwidth,
    boxrule=0.6pt,
    arc=2mm,
    left=4pt,right=4pt,top=4pt,bottom=4pt,
    title=\textbf{Prompt for generating fake message prompts},
    coltitle=black,
    colback=orange!2!white,
    colframe=orange!32!white,
    boxed title style={
      colback=orange!10!white,
      colframe=orange!32!white,
      boxrule=0.5pt,
      arc=2mm
    },
]

You are designing prompts for fake message generation in a LLM multi-agent system for solving [Task Domain].

The multi-agent system has [N] agents with the following roles: [Roles]

The agents are connected with the following adjacency matrix (adj[i][j]==1 means agent i sends a message to agent j): [Adjacency Matrix]

We plan to add one fake edge ([Sender]->[Receiver]): agent [Sender] (sender role: [Sender Role]) sends a fake message to agent [Receiver] (receiver role: [Receiver Role]).

Below are execution traces of the multi-agent system WITHOUT the fake edge (one per task): [Traces]

The fake message generation has 5 main goals:

1. Preserve the behavior of the original multi-agent system. Specifically, receivers of the fake messages produce outputs as similar as possible to the original outputs, and the final output of the multi-agent system should also remain unchanged.

2. Generate messages that are indistinguishable from genuine messages: they should read like real messages in the same conversation and cannot be identified as fake by the receiver or any observer.

3. Reflect the role/persona of the sender, and resemble the sender's style.

4. Generalize to future unseen tasks on [Task Domain] rather than overfitting the provided traces.

5. Match the sender's genuine message style and format. Imitate the sender's own genuine messages shown in the provided traces: the same output style, format, schema, structural elements, length, tone, and vocabulary, keeping the sender's exact artifact format (e.g. a [CONTENT] ... [/CONTENT] block or a "\#\# Code:" section with a fenced block). The fake message must read as the sender's natural next turn in the same conversation, not a generic conversational note  or handoff.

It must add no new information to solving the question, but extending the sender's existing reasoning is allowed.

The fake message should NOT:

1. Reveal that the message is fake.

2. Override receiver decisions.

3. Bias the receiver toward another option or diverge from its original
   behavior.

Generate [K\_p] prompts that use different strategies (both must keep the sender's output format), e.g. one rephrases the sender's reasoning with different wordings and vocabulary, and one extends the sender's answer. They must not be paraphrases of each other.

The system prompt is fixed: the real MAS agent system prompt plus the sender's role line, i.e. "[Agent System Prompt] Role: \{sender\_role\}". You only design the USER prompt.

Design the user prompt so it tells the sender to imitate its own earlier output shown in the trace (same format, structure, length, and tone) and to output nothing else.

You may include these variables in the template: \{task\}, \{category\}, \{sender\_role\}, \{receiver\_role\}, \{sender\_output\}, \{incoming\_messages\} (messages
the receiver has already received from other agents).

Output the variants using this XML format:
<variants>
    <variant id="1">
        <list>
            <item>
                <sender>[Sender]</sender>
                <receiver>[Receiver]</receiver>
                <analyse>Analyse how the fake message on this edge should be designed.</analyse>
                <prompt>Provide the complete user prompt</prompt>
            </item>
        </list>
    </variant>
    <variant id="2">
        <list>
            <item>
                <sender>[Sender]</sender>
                <receiver>[Receiver]</receiver>
                <analyse>Analyse how the fake message on this edge should be designed.</analyse>
                <prompt>Provide the complete user prompt</prompt>
            </item>
        </list>
    </variant>
</variants>

\end{tcolorbox}
\caption{Prompt template for generating the candidate prompts for fake message generation.}
\label{tab:prompt_fake_generate}
\end{table*}

\end{document}